\documentclass[12pt, fleqn, leqno]{article}
\usepackage{amsmath}\usepackage{amsfonts}
\usepackage{amsbsy}
\usepackage{latexsym}
\mathindent=.6in\numberwithin{equation}{section}

\newcommand{\be}{\begin{equation}}
\newcommand{\ee}{\end{equation}}
\newcommand{\ba}{\begin{array}}
\newcommand{\ea}{\end{array}}

\renewcommand{\v}{{\mathsf{v}}}

\newcommand{\bt}{\beta}

\renewcommand{\b}{\beta}

\newcommand{\bea}{\begin{eqnarray}}
\newcommand{\eea}{\end{eqnarray}}

\newcommand{\Ga}{\Gamma}

\newcommand{\al}{\alpha}
\newcommand{\pri}{\prime}

\begin{document}
\newtheorem{pro}[thm]{Proposition}
\newtheorem{lem}[thm]{Lemma}
\newtheorem{cor}[thm]{Corollary}
\newcommand{\C}{{\mathsf{C}}}
\newcommand{\A}{{\mathsf{A}}}
\newcommand{\G}{{\mathsf{\Gamma}}}
\renewcommand{\a}{{\mathsf{a}}}
\title{Perturbed Hankel Determinants.}
\author{Estelle Basor$^{\dag}$\\
Department of Mathematics\\
Calpoly, San Luis Obispo, USA\\
Yang Chen$^{*}$\\
Center for Combinatorics\\
Nankai University\\
Tianjin, China}
\date{}\maketitle
\begin{abstract} 
\noindent
In this short note, we compute, for large $n$ the determinant of a class 
of $n\times n$ Hankel matrices, which arise from a smooth perturbation of the 
Jacobi weight. For this purpose, we employ the same idea used
in previous papers, where the unknown determinant,
$D_n[w_{\al,\bt}h]$ is compared with the known determinant 
$D_n[w_{\al,\bt}].$ Here $w_{\al,\bt}$ is the Jacobi weight and
$w_{\al,\bt}h,$ where $h=h(x),\;x\in[-1,1]$ is strictly positive and real
analytic, is the smooth perturbation on the Jacobi weight 
$w_{\al,\bt}(x):=(1-x)^{\al}(1+x)^{\bt}.$
Applying a previously known formula on the distribution function of 
linear statistics,
we compute the large $n$ asymptotics of $D_n[w_{\al,\bt}h]$ and 
supply a missing constant of the expansion.
\end{abstract}

{\bf {Running title}}: Hankel Determinants, perturbations.
\\
$^{\dag}$ebasor@calpoly.edu. Supported in part by NSF Grants DMS-0200167 and
DMS-0500892. \\
$^*$ychen@ic.ac.uk. Supported in part by EPSRC Grant EP/C534409/01.\\ 
On leave from Imperial College London.
\newpage 
\section{Introduction and Preliminaries.}

The purpose of this note is to find heuristically an asymptotic 
expansion for determinants of certain Hankel matrices. The matrices are 
generated by the moments of a function defined on the interval $[-1,1].$ 
Let $w(x)$ be a function of the form
$$ w_{\alpha,\beta}(x)h(x) $$
where
$$w_{\alpha, \beta}(x) = (1-x)^{\alpha}(1+x)^{\beta},\quad\al\geq 0,
\;\bt\geq 0$$
and 
$h(x)$ is a strictly positive function with a derivative satisfying 
a Lipschitz condition. 

Define 
$$\mu_{k}[w]=\int_{-1}^{1}x^{k}w(x)dx,\quad k=0,1,2,..$$
and
$$
D_{n}[w] = \det(\mu_{j+k}[w])_{j,k=0}^{n-1}.
$$
The motivation for investigating such perturbed Hankel 
determinants, comes from
Random Matrix Theory and its applications, where one studies the
generating functions of linear statistics \cite{CM,CL}. 
Also see \cite{BCW1} and some of the references  
in that volume.   

 Our goal will be to show formally that, for $w=w_{\al,\bt}h,$ 
\be
D_{n}[w] \sim 2^{-n(n+\al+\bt)}n^{(\al^2+\bt^2)/2-1/4}(2\pi)^n
\exp\left(\frac{n}{\pi} \int_{-1}^{1}
\frac{\ln h(x)}{{\sqrt {1-x^2}}}dx\right)C
\ee
where the $n$ independent constant $C$ is given by
$$
\exp\left[\frac{1}{4\pi^2}\int_{-1}^{1}
\frac{\ln h(x)}{{\sqrt {1-x^2}}}
\left(P\int_{-1}^{1}
\frac{{\sqrt {1-y^2}}}{y-x}
\frac{h^{\prime}(y)}{h(y)}dy\right)dx \right]
$$
$$ 
\times \exp\left( \frac{\al+\bt}{2 \pi}\int_{-1}^{1}
\frac{\ln h(x)}{{\sqrt {1-x^2}}}dx 
\right)\frac{G^2\left(\frac{\al+\bt+1}{2}\right)
G^2\left(\frac{\al+\bt}{2}+1\right)\Ga\left(\frac{\al+\bt+1}{2}\right)}
{G(\al+\bt+1)G(\al+1)G(\bt+1)}.
$$
In the above formula the function $G$ is the Barnes $G$-function, 
an entire function that satisfies the difference equation 
$G(z+1)=\Ga(z)G(z),$ with $G(1)=1.$ The result (1.1) is also valid for
$\al\geq -1/2,$ and $\bt\geq -1/2,$ since this expression is real analytic
in $\al$ and $\bt.$

The main idea, which can be traced back to a paper of 
Szeg\H{o} \cite{Sze1}, is that one can find the above formula in two steps. 
The first is to consider 
the ``pure'' weight 
$$w_{\alpha, \beta}(x) = (1-x)^{\alpha}(1+x)^{\beta}.$$
Using some basic results from the theory of orthogonal 
polynomials the Hankel determinant for 
the pure weight can be found exactly and then easily computed asymptotically. 
This step is rigorous and in fact may be the first 
instance where these asymptotics are found completely.

The next step is to use the linear statistics formula derived
from the Coulomb fluid approach \cite{CM,CL}-- expected to be valid for 
sufficiently large $n$--
to compute the quotient
\be \frac{D_{n}[w_{\al,\bt}h]}{ D_{n}[w_{\al, \beta}]} ,\ee
thus achieving the desired result.

We note that in a recent work \cite{RH}, 
the asymptotic formula for $D_{n}$ appears, but without the constant term. 
In future work, we hope to use 
the techniques of \cite{BCW2} to make the 
ideas presented here complete and thus firmly establish the 
validity of the asymptotic formula.
 
 We begin with some notation. Let $P_n(x)$ be monic polynomials 
of degree $n$ in $x$ and orthogonal, with respect to a weight, 
$w(x),\;x\in[a,b];$ 
\bea\int_{a}^{b}P_m(x)P_n(x)w(x)dx=h_n[w]\delta_{m,n},
\eea
where $h_j[w]$ is the square of the $L^2$ norm of the
polynomials orthogonal with respect to $w,$ over $[-1,1].$ 

 From the orthogonality condition there follows the recurrence relation,
\bea zP_n(z)=P_{n+1}(z)+\al_nP_n(z)+\bt_nP_{n-1}(z),
n=0,1,...,\;
\eea
where $\bt_0 P_{-1}(z):=0$, $\al_n,\;n=0,1,2,...$ is real and
$\bt_n>0, \;n=1,2,...$ 

There is an intimate relationship between the values of 
$\beta_{n}$, $h_n$ and the Hankel determinants.

 Indeed, the determinant, for any weight $w,$ 
$$D_{n}[w] = \prod_{j=0}^{n-1}h_j[w]. $$

 In addition, 
 $$h_j[w] =h_0[w] \prod_{k =1}^{j} \beta_{k}.$$
 Thus if we can compute $\beta_{j}$ it follows that both
 $h_{j}$ and $D_{n}$ can be explicitly determined.
 For this and all other basic results see \cite{Sz}. 
 
 For the monic Jacobi polynomials, that is in the case when
 $w = w_{\al, \beta},$ 
it is well-known that
$$
\al_n=\frac{\bt^2-\al^2}{(2n+\al+\bt)(2n+\al+\bt+2)}
$$
 and
$$
\bt_n=\frac{4n(n+\al)(n+\bt)(n+\al+\bt)}{(2n+\al+\bt)^2(2n+\al+\bt+1)
(2n+\al+\bt-1)}.
$$
 Hence it follows that 
\be 
h_n[w_{\al,\bt}]
 \ee
 $$ 
 = 2^{2n+\al+\bt+1} \frac{\Ga(n+1)\Ga{(n+\al+1)}\Ga(n+\bt+1)
\Ga(n+\al+\bt+1)}{(2n+\al+\bt+1)
[\Ga(2n+\al+\bt+1)]^2},$$
\newpage 
and
\be
D_n[w_{\al,\bt}]
\ee
$$= 2^{-n(n+\al+\bt)}(2\pi)^n
\frac{\Ga\left(\frac{\al+\bt+1}{2}\right)G^2\left(\frac{\al+\bt+1}{2}
\right)G^2\left(\frac{\al+\bt}{2}
+1\right)}{G(\al+\bt+1)G(\al+1)G(\bt+1)}$$
$$
\times \frac{G(n+1)G(n+\al+1)G(n+\bt+1)G(n+\al+\bt+1)}
{G^2\left(n+\frac{\al+\bt+1}{2}\right)
G^2\left(n+\frac{\al+\bt}{2}+1\right)
\Ga\left(n+\frac{\al+\bt+1}{2}\right)}
$$
where $G(z)$ is the Barnes $G$-function.
 See \cite{Chen} for a first-principle derivation of the recurrence 
coefficients.

The asymptotics of the Gamma function and the Barnes $G-$function are well 
understood. We have  that
 \bea
\mbox{} \Ga(n+a)&\sim& {\sqrt {2\pi}}{\rm e}^{-n}n^{n+a-1/2},\nonumber\\
\mbox{} G(n+a+1)&\sim& n^{(n+a)^2/2-1/12}{\rm e}^{-3n^2/4-an}(2\pi)^{(n+a)/2}K,
\nonumber\\
\mbox{}{\rm where\;\;}K&:=&G^{2/3}(1/2)\pi^{1/6}2^{-1/36}.\nonumber
\eea
>From the above asymptotic expressions an easy computation shows that,
\bea
\mbox{} D_{n}[w_{\al,\bt}]&\sim& 2^{-n(n+\al+\bt)}n^{(\al^2+\bt^2)/2-1/4}
(2\pi)^n\nonumber\\
&\times&\frac{G^2\left(\frac{\al+\bt+1}{2}
\right)G^2\left(\frac{\al+\bt}{2}+1\right)
\Ga\left(\frac{\al+\bt+1}{2}\right)}
{G(\al+\bt+1)G(\al+1)G(\bt+1)}.
\eea
 The above formula is the promised result for the ``pure''  weight.
When $\al=0=\bt,$ we find,
$$
D_n[w_{0,0}]\sim \frac{\pi^n}{n^{1/4}}\frac{1}{2^{n(n-1)}}
\;G^2(1/2)\Gamma(1/2).
$$
This is consistent with Hilbert's \cite{Hilbert} 
asymptotic expression for large $n,$ of the Hankel determinant associated
with the Legendre weight,
$$
\left(D_n[w_{0,0}]\right)^{1/n}
=\frac{\pi}{2^{n-1}}\left(1+\varepsilon_n\right),\;\;\;\;{\rm where\;\;}
\lim_{n\to\infty}\varepsilon_n=0.
$$
and we have changed the notations of \cite{Hilbert} to be compatible 
with ours. 
\setcounter{thm}{0}
\setcounter{equation}{0}
\section{Perturbed Jacobi Weight}
\setcounter{section}{2}
In this section we  show how to compare the unknown Hankel 
determinant $D_n[w_{\al,\bt}h],$ with the known Hankel determinant 
$D_n[w_{\al,\bt}].$
It is known from \cite{Sz} (see also \cite{CM,CL}) that
\bea
\frac{D_n[w_{\al,\bt}h]}{D_n[w_{\al,\bt}]}=\left<\prod_{k=1}^{n}h(x_j)\right>,
\eea
where ${\cal \Psi}={\cal \Psi}(x_1,...,x_n),$ and 
 \bea
\left<{\cal \Psi}\right>:=\frac{\int_{-1}^{1}...\int_{-1}^{1}
{\cal \Psi}
\prod_{1\leq j<k\leq n}(x_k-x_j)^2\prod_{l=1}^nw_{\al,\bt}(x_l)dx_l}
{\int_{-1}^{1}...\int_{-1}^{1}\prod_{1\leq j<k\leq n}(x_k-x_j)^2
\prod_{l=1}^nw_{\al,\bt}(x_l)dx_l}.
\eea

 This can be rewritten  as an average of the exponential of the linear 
statistics $\sum_{l=1}^{n}\ln h(x_l),$ i.e.,
$$
\left<\exp\left(\sum_{l=1}^{n}\ln h(x_l)\right)\right>.
$$
 Note, because of the assumptions on $h,$ $\ln h$ is well defined 
for $x\in [-1,1].$ 
 Results, at least in a heuristic way, are known about such 
linear statistics. In particular, the logarithm of (2.1) is, for large $n$,
$$
\frac{1}{4\pi^2}\int_{a_n}^{b_n}
\frac{\ln h(x)}{{\sqrt {(b_n-x)(x-a_n)}}}
\left(P\int_{a_n}^{b_n}\frac{{\sqrt {(b_n-y)(y-a_n)}}}{y-x}
\frac{h^{\prime}(y)}{h(y)}dy\right)dx
$$
$$
+\int_{a_n}^{b_n}\ln h(x)\sigma(x)dx,\eqno(2.3)
$$
where the equilibrium density $\sigma(x)$, defined for
$x\in[a_n,b_n]$ is
$$
\sigma(x)=\frac{\sqrt {(b_n-x)(x-a_n)}}{2\pi^2}
\int_{a_n}^{b_n}\frac{\v^{\prime}(x)-\v^{\prime}(y)}{x-y}
\frac{dy}{{\sqrt {(b_n-y)(y-a_n)}}},
$$
and 
$$\v^{\pri}(x):=-\frac{w_{\al,\bt}^{\prime}(x)}{w_{\al,\bt}(x)}
=-\frac{\al}{x-1}-\frac{\bt}{x+1}.$$ 
The end points $a_n,\;b_n,$ of the support 
are determined by,
\bea 
\mbox{} 2\pi n&=&\int_{a_n}^{b_n}\frac{x\v^{\prime}(x)}
{{\sqrt {(b_n-x)(x-a_n)}}}dx,\nonumber\\
\mbox{} 0&=&\int_{a_n}^{b_n}\frac{\v^{\prime}(x)}{{\sqrt {(b_n-x)(x-a_n)}}}dx.
\nonumber
\eea
The equation (2.3) was derived in \cite{CM} and later a large $n$ version
was found in \cite{CL}. For another derivation of (2.3) 
using a ``small fluctuations'' approach see \cite{BCW1}.

In our problem, the above equations become,
\bea
\mbox{} n+\left(\frac{\al+\bt}{2}\right)&=&
\frac{\al}{2{\sqrt {(1-a_n)(1-b_n)}}}
+\frac{\bt}{2{\sqrt {(1+a_n)(1+b_n)}}}\nonumber\\
\mbox{} 0&=&\frac{\al}{{\sqrt {(1-a_n)(1-b_n)}}}
-\frac{\bt}{{\sqrt {(1+a_n)(1+b_n)}}},\nonumber
\eea
and the solutions are
\bea
\mbox{} a_n&=&\frac{\bt^2-\al^2-4{\sqrt {n(n+\al)(n+\bt)(n+\al+\bt)}}}
{(2n+\al+\bt+2)^2}\nonumber\\
\mbox{} b_n&=&\frac{\bt^2-\al^2+4{\sqrt {n(n+\al)(n+\bt)(n+\al+\bt)}}}
{(2n+\al+\bt+2)^2}.\nonumber
\eea
In the Coulomb fluid approximations \cite{chen-coul}, 
the diagonal ($\tilde{\al}_n$) and off-diagonal 
recurrence coefficients ($\tilde{\bt}_{n}$) are
\bea
\mbox{} \tilde{\al}_n&=&\frac{b_n+a_n}{2}=\frac{\b^2-\al^2}{(2n+\al+\bt)^2},
\nonumber\\
\mbox{} \tilde{\bt}_n&=&\frac{(b_n-a_n)^2}{16}
=\frac{4n(n+\al)(n+\bt)(n+\al+\bt)}{(2n+\al+\bt)^4},\nonumber
\eea
and the deviations from the exact results are,
\bea
\mbox{} \tilde{\al}_n-\al_n&=&\frac{\bt^2-\al^2}{4n^3}+
{\rm O}\left(\frac{1}{n^4}\right),\nonumber\\
\mbox{} \tilde{\bt}_n-\bt_n&=&-\frac{1}{16n^2}+
{\rm O}\left(\frac{1}{n^3}\right).\nonumber
\eea
For later reference we also note that,
\bea
\mbox{} 1+a_n&=&\frac{\bt^2}{2n^2}+{\rm O}\left(\frac{1}{n^3}\right)
\nonumber\\
\mbox{} 1-b_n&=&\frac{\al^2}{2n^2}+{\rm O}\left(\frac{1}{n^3}\right).
\nonumber
\eea 
A simple calculation shows that, for $x\in [a_n,b_n],$ 
\bea 
\mbox{} &&\frac{\sigma(x)}{\sqrt {(b_n-x)(x-a_n)}}\nonumber\\
&=&\frac{1}{2\pi}
\left[\frac{\al}{{\sqrt {(1-a_n)(1-b_n)}}(1-x)}+
\frac{\bt}{{\sqrt {(1+a_n)(1+b_n)}}(1+x)}\right]\nonumber\\
\mbox{} &=&\left(n+\frac{\al+\bt}{2}\right)
\frac{1}{1-x^2},\;\;\;-1< a_n<b_n<1.
\nonumber
\eea
where we have used,
\bea
\mbox{} 
\frac{\al}{{\sqrt {(1-a_n)(1-b_n)}}}&=&n+\frac{\al+\bt}{2},\nonumber\\
\mbox{} 
\frac{\bt}{{\sqrt {(1+a_n)(1+b_n)}}}&=&n+\frac{\al+\bt}{2}.\nonumber
\eea
Therefore, for $x\in(-1,1),$ and $n$ large,
\bea
\sigma(x)=\frac{n+(\al+\bt)/2}{\pi{\sqrt {1-x^2}}}+
{\rm O}\left(\frac{1}{n}\right).\nonumber
\eea
Put $f(x)=\ln h(x),$ and $x=R_n+ r_nt,$ where $R_n:=(b_n+a_n)/2,$ and 
$r_n:=(b_n-a_n)/2,$ the second term of (2.3) becomes 
\bea
\left(n+\frac{\al+\bt}{2}\right)
r_n^2\int_{-1}^{1}\frac{f(R_n+r_nt)}{1-(R_n+r_nt)^2}{\sqrt {1-t^2}}dt,
\nonumber
\eea
while the first term of (2.3) reads,
\bea
\frac{r_n}{4\pi^2}\int_{-1}^{1}\frac{f(R_n+r_ns)}{\sqrt {1-s^2}}
\left(P\int_{-1}^{1}\frac{{\sqrt {1-t^2}}}{t-s}f^{\prime}(R_n+r_nt)dt\right)ds.
\nonumber
\eea
Now, since 
\bea
\mbox{}R_n&=&\frac{\bt^2-\al^2}{4}\frac{1}{n^2}
+{\rm O}\left(\frac{1}{n^3}\right)\nonumber\\
\mbox{}r_n&=&1-\frac{\al^2+\bt^2}{4}\frac{1}{n^2}+
{\rm O}\left(\frac{1}{n^3}\right)\nonumber
\eea
we see that the second term of (2.3) is asymptotic to
\bea
\left(n+\frac{\al+\bt}{2}\right)\int_{-1}^{1}\frac{\ln h(x)}
{\pi{\sqrt {1-x^2}}}dx+{\rm o}(1),\nonumber
\eea
while the first term of (2.3) is asymptotic to
\bea
\frac{1}{4\pi^2}\int_{-1}^{1}\frac{\ln h(x)}{{\sqrt {1-x^2}}}
\left(P\int_{-1}^{1}\frac{{\sqrt {1-y^2}}}{y-x}
\frac{h^{\prime}(y)}{h(y)}dy\right)dx+{\rm o}(1).\nonumber
\eea
The above two expressions combined with (1.7) give the formula (1.1).

\end{document}